\documentclass[conference]{IEEEtran}
\IEEEoverridecommandlockouts
\usepackage{cite}
\usepackage{amsmath,amssymb,amsfonts}
\usepackage{graphicx}
\usepackage{textcomp}
\usepackage{xcolor}
\usepackage{bm}
\usepackage{graphicx}
\usepackage{xcolor}
\usepackage{subcaption}
\usepackage{amsthm}
\usepackage{graphicx} %导入图片需要的宏包
\usepackage{epsfig}%.eps数据图片 这一句一定要加上 
\usepackage{epstopdf}%.eps数据图片
\usepackage{booktabs}  % 用于美观的表格
\usepackage{graphicx}
\usepackage{subcaption}
\usepackage{mathtools} % 加载 mathtools 宏包
\usepackage{soul} 
\usepackage{algorithm}
\usepackage[top=0.7in, bottom=0.99in, left=0.625in, right=0.625in]{geometry}
\usepackage[noend]{algpseudocode} % 使用 noend 选项避免 \End 问题
\makeatletter
\@ifundefined{thealgorithm}{}{} % 保存原有定义
\makeatother

\algnewcommand{\Input}{\item[\textbf{Input:}]}
\algrenewcommand{\Ensure}{\item[\textbf{Output:}]}

\def\BibTeX{{\rm B\kern-.05em{\sc i\kern-.025em b}\kern-.08em
		T\kern-.1667em\lower.7ex\hbox{E}\kern-.125emX}}

\begin{document}

\title{Timeliness-Aware Joint Source and Channel Coding for Adaptive Image Transmission}

\author{
    \IEEEauthorblockN{
        Xiaolei Yang,~\IEEEmembership{Student Member,~IEEE}, 
        Zijing Wang,~\IEEEmembership{Member,~IEEE},
        Zhijin Qin,~\IEEEmembership{Senior Member,~IEEE}, \\
        Xiaoming Tao,~\IEEEmembership{Senior Member,~IEEE}
    }\\
    %\textsuperscript{1}School of Electronic Information Engineering, Beijing Jiaotong University, China \\
    Department of Electronic Engineering, Tsinghua University, Beijing, China\\
    State Key Laboratory of Space Network and Communications, Beijing, China \\
    Beijing National Research Center for Information Science and Technology, Beijing, China
    \thanks{This work is supported in part by the National Key Research and Development Program of China under Grant 2023YFB2904300, in part by the National Natural Science Foundation of China (NSFC) under Grant 62293484 and Grant 62401321, and in part by the China Postdoctoral Science Foundation under Grant 2024M751681.}
}

\maketitle

\begin{abstract}

Accurate and timely image transmission is critical for emerging time-sensitive applications such as remote sensing in satellite-assisted Internet of Things. However, the bandwidth limitation poses a significant challenge in existing wireless systems, making it difficult to fulfill the requirements of both high-fidelity and low-latency image transmission. Semantic communication is expected to break through the performance bottleneck by focusing on the transmission of goal-oriented semantic information rather than raw data. In this paper, we employ a new timeliness metric named the value of information (VoI) and propose an adaptive joint source and channel coding (JSCC) method for image transmission that simultaneously considers both reconstruction quality and timeliness. Specifically, we first design a JSCC framework for image transmission with adaptive code length. Next, we formulate a VoI maximization problem by optimizing the transmission code length of the adaptive JSCC under the reconstruction quality constraint. Then, a deep reinforcement learning-based algorithm is proposed to solve the optimization problem efficiently. Experimental results show that the proposed method significantly outperforms baseline schemes in terms of reconstruction quality and timeliness, particularly in low signal-to-noise ratio conditions, offering a promising solution for efficient and robust image transmission in time-sensitive wireless networks.

\end{abstract}

% \begin{IEEEkeywords}
% component, formatting, style, styling, insert
% \end{IEEEkeywords}

% \section{Introduction}

\section{Introduction}
High fidelity and timeliness are very important in image reconstruction tasks for many emerging wireless applications, such as remote sensing and monitoring in satellite networks and Internet of Things (IoT) networks. However, these applications face the bandwidth bottleneck challenge, which is unable to support high-quality and timely image transmission in the low signal-to-noise ratio condition. Semantic communication has offered a promising alternative, aiming to transmit goal-oriented information rather than raw data to enable more intelligent and efficient communication in the 6G era\cite{qin2024ai,tensemantic}.

The deep joint source-channel coding (JSCC) is a key technique in semantic communication to enable high-quality data reconstruction under weak channel conditions, which breaks the separation in traditional wireless networks by designing source and channel coding jointly. Bourtsoulatze \textit{et al.} \cite{bourtsoulatze2019deep} first proposed a deep JSCC method using convolutional neural networks (CNN) for wireless image transmission, which illustrated great performance gain of joint coding compared with traditional separate counterpart, especially under low signal-to-noise ratio (SNR) and limited bandwidth conditions. However, the fixed networks architecture of deep JSCC constrains its adaptability to varying channel conditions and different coding rates. Considering the limitations of deep JSCC, a variety of adaptive JSCC architectures \cite{xu2021wireless} \cite{bian2023deepjscc} have been developed to dynamically adjust to varying channel SNRs and transmission code lengths. Moreover, several studies \cite{yang2024swinjscc} \cite{dai2022nonlinear} have explored the use of advanced neural architectures to further enhance the performance and flexibility of JSCC systems. 

Despite their contributions, timeliness has not been explicitly studied. For example, in real-time IoT monitoring systems, the effectiveness of image transmission depends not only on the data's content but also on the timeliness of the transmitted information. To quantify the temporal level of semantic information, age of information (AoI) was first introduced as a metric to measure data freshness, effectively reflecting the end-to-end delay of information updates \cite{yates2021age}. Beyond AoI, several enhanced metrics \cite{maatouk2022age,li2024goal,agheli2022semantics,wang2022framework} have been proposed to measure both freshness and other semantic characteristics, such as the correctness of the received data \cite{maatouk2022age} and the semantic mismatches given an underlying task \cite{li2024goal}. They are regarded as a useful optimization tool in the design of timeliness-aware applications. Particularly, the value of information (VoI) was introduced in the context of information theory \cite{wang2022framework}, which incorporates freshness, source dynamics, and channel conditions, providing a comprehensive characterization of timeliness.

In current deep JSCC systems, the adaptability to different coding rates lacks the consideration of information timeliness. It has been empirically observed that longer codewords in JSCC lead to improved reconstruction quality, while simultaneously causing higher transmission latency. This trade-off can be particularly problematic in time-sensitive bandwidth-constrained scenarios such as remote IoT and emergency communications, where timely updates are essential for effective decision-making and emergency response. Therefore, it is essential to consider both content-level semantics and temporal-level semantics in the system design. 

Motivated by the above observations, we propose a timeliness-aware semantic communication framework that jointly considers reconstruction quality and transmission timeliness, using an adaptive semantic codec and dynamic code length allocation. The main contributions of this paper are summarized as follows:

\begin{itemize}

\item  We propose a novel JSCC framework for image transmission with adaptive code length, which is able to simultaneously guarantee the reconstruction quality and timeliness.

\item We utilize a new performance metric and formulate an optimization problem to enhance information timeliness by controlling the code length of JSCC under reconstruction quality constraints.

\item We implement an adaptive JSCC codec along with a deep reinforcement learning (DRL)-based optimizer, achieving an effective trade-off between image reconstruction quality and information timeliness.

\end{itemize}

\section{System model}

\begin{figure}[t]
\centering 
\includegraphics[width=\linewidth]{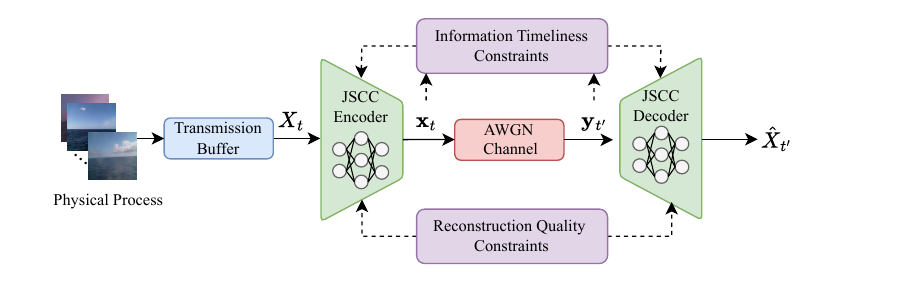}
\caption{Framework of remote IoT monitoring system.}\vspace{-0.3cm}
\label{fig:system_model}
\vspace{-0.4cm}
\end{figure}

We consider a timeliness-aware semantic image transmission system as shown in Fig.~\ref{fig:system_model}. An IoT device periodically captures images from the physical process and transmits them over a wireless channel to a remote receiver.

%\subsection{Adaptive Image Transmission}
At the time $t$, the transmission buffer at the transmitter only holds a single image, denoted as $X_t \in \mathbb{R}^{C \times H \times W}$, which corresponds to the most recently sampled observation from the physical environment. This image is generated periodically with a fixed sampling interval $T_s$, and the $n$-th image's generation timestamp is defined as $u(t) = n \cdot T_s$. Older images are discarded immediately upon the arrival of a new sample, ensuring that the system always processes the freshest available data.

The wireless channel is intermittently available. If the channel is currently busy, the system waits until the channel becomes available before initiating the transmission. When the channel becomes idle, the transmitter encodes the buffered image $X_t$ into the complex semantic code $ \mathbf{x}_t \in \mathbb{C}^{K} $ as
\begin{equation}
    \mathbf{x}_t = f_{\text{enc}}(X_t, \gamma_t, \eta_t ;\Theta_{\text{enc}}),
\end{equation}
where $f_{\text{enc}}$ is the encoder function, $\Theta_{\text{enc}}$ represents the corresponding network parameters, $\gamma_t$ is current channel SNR, $\eta_t$ represents current channel bandwidth ratio (CBR) and $\eta = \frac{K}{C \times H \times W}$. The encoder’s adaptability to channel SNR and CBR enhances its robustness against channel variations and enables dynamic adjustment to meet timeliness requirements for transmission.

The encoded codeword $\mathbf{x}_t$ is transmitted over an additive white Gaussian noise (AWGN) channel, and the received signal at the receiver is given as
\begin{equation}
    \mathbf{y}_{t'} = \mathbf{x}_t + \mathbf{n}_{t'}, 
\end{equation}
where ${t'}$ is the received time stamp, $\mathbf{n}_{t'}$ denotes the Gaussian noise vector and $\mathbf{n}_{t'} \sim \mathcal{N}(0, \sigma_{t'}^2 \mathbf{I})$, and $\sigma_{t'}^2$ represents the noise variance at time $t'$. 
% The received signal $\mathbf{y}_{t'}$ may be corrupted due to channel noise, which motivates the use of error control mechanisms.
The receiver will reconstruct the image using a JSCC decoder as
\begin{equation}
    \hat{X}_{t'} = f_{\text{dec}}(\mathbf{y}_{t'}; \Theta_{\text{dec}}),
\end{equation}
where $f_{\text{dec}}$ is the decoder function, and $\Theta_{\text{dec}}$ represents the corresponding network parameters. 

In Fig.~\ref{fig:system_model}, the reconstruction quality is measured by the peak signal-to-noise ratio (PSNR) which is given as 
 \begin{multline}
     \text{PSNR}(X_t, \hat{X}_{t'}) \\
     = 10 \cdot \log_{10} \left( \frac{MAX^2}{\frac{1}{HWC} \sum_{i,j,k} (X(i,j,k) - \hat{X}(i,j,k))^2} \right),
 \end{multline}
 where $MAX$ is the maximum possible pixel value (e.g., 255), and $H$, $W$, and $C$ denote the height, width, and number of channels of the image, respectively.

%\subsection{Timeliness Performance Metric}
The timeliness of information is measured by the VoI \cite{wang2022framework}. The VoI is initially defined as the mutual information between the current status of the monitoring process and the latest noisy observation at the receiver, which is given as
\begin{equation}
    \text{VoI}(t) = \mathbb{I}(\mathcal{X}_t; \mathcal{\hat{X}}_{t'}).
\end{equation}
Here, $\mathcal{X}_t$ and $\mathcal{\hat{X}}_{t'}$ denote the underlying temporal-level semantic significance of the original image $X_t$ and the reconstructed image $\hat{X}_{t'}$. The temporal-level semantic significance of the image can be seen as a random process. We assume the random process is stationary, Markovian, and Gaussian. This is a common simplification that makes the problem analytically tractable, allowing us to derive a closed-form expression for the VoI as:
\begin{equation}
    \text{VoI}(t) = - \log(1- \frac{\gamma_t}{1+\gamma_t}\rho^{A(t)}),
\end{equation}
where $\rho$ is the autocorrelation coefficient between the different time-relevant semantic significance of two successive images. $A(t)$ represents the AoI, which is a widely used metric to quantify timeliness which measures the time elapsed since the generation of the most recently received update. Specifically, the AoI is defined as ${A}(t)=t-u(t)$. 

% \textcolor{blue}{It could be added a short paragraph to generally introduce the relationship of works in following sections.}
The rest of this paper is organized as follows. Section III presents the proposed adaptive JSCC framework. Section IV formulates the timeliness-aware transmission optimization problem and introduces a DRL-based codeword length allocator. Section V shows experimental results, and Section VI concludes the paper.
% to the application task. In practical semantic communication systems, however, the importance of information varies dynamically.

% The value of information (VoI) quantifies the relevance and importance of the received data, taking into account both its freshness and its semantic content.

\begin{figure*}[t]
    \centering
    % \vspace{-0.5cm}
    \includegraphics[width=0.95\textwidth]{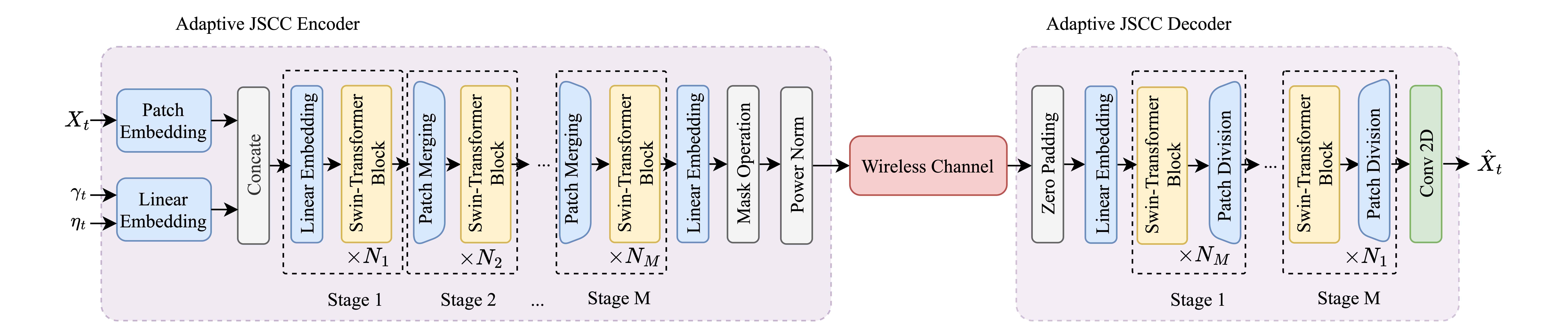}
    \caption{Architecture of proposed JSCC for adaptive image transmission.}
    \label{fig:jsccnet} 
    \vspace{-0.6cm}
\end{figure*}

\begin{figure}[t]
    \centering
    % \vspace{-0.5cm}
    \includegraphics[width=0.48\textwidth]{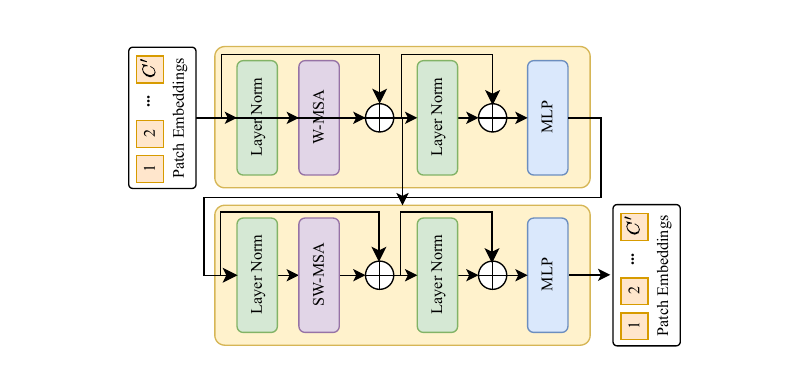}\vspace{-0.3cm}
    \caption{Details of two successive Swin Transformer blocks.}
    \label{fig:swin} 
    \vspace{-0.6cm}
\end{figure}

\section{Adaptive Joint Source-Channel Coding}
The proposed adaptive joint source and channel coding framework can be seen as Fig.~\ref{fig:jsccnet}. 
The input image $X_t \in \mathbb{R}^{H\times W \times C}$ is first divided into $\frac{H}{2} \times \frac{W}{2}$ non-overlapping patches and passed through a linear embedding layer, which maps each patch into a latent representation. 
In parallel, both $\gamma_t$ and $\eta_t$ are projected via linear layers to form side-information embeddings.
Inspired by the related work \cite{bian2023deepjscc}, we also repeated the side information embeddings to match the number of image patches and then concatenated with each corresponding patch embedding along the feature dimension. The concatenated tokens are then fed into $N_1$ linear embedding layers and Swin Transformer blocks. The details of two successive Swin Transformer blocks are shown in Fig.~\ref{fig:swin}. The first block utilizes the window-based multi-head self-attention (W-MSA) mechanism to capture local dependencies within each patch, while the second block employs the shifted window-based multi-head self-attention (SW-MSA) mechanism to capture global dependencies across patches. This hierarchical structure allows the model to effectively learn both local and global features from the input image \cite{liu2021swin}. In order to enhance the hierarchical representation learning ability, the output tokens are fed into the patch merging layer, which reduces the spatial resolution of the feature map by a factor of 2 while increasing the channel dimension. This process enables the model to capture more abstract and high-level features in the subsequent layers. And several Swin Transformer blocks are stacked to further enhance the feature representation capability of the model. In order to achieve the adaptive code length control, a masking operation is introduced after the final block. Only the first few tokens are retained based on the CBR, and the selected tokens are then passed through a power normalization module to satisfy the average transmit power constraint, producing the final encoded codeword $\mathbf{x}_t$.

At the receiver, the noisy codeword $\mathbf{y}_{t'}$ is received through an AWGN channel. Since the decoder expects a fixed-length latent representation, the received codeword is first zero-padded to the maximum length before being passed through the decoder. The decoder mirrors the encoder structure and reconstructs the output image $\hat{X}_{t'}$ from the padded latent representation.

To optimize the reconstruction quality, the entire JSCC framework is trained in an end-to-end manner using the mean squared error (MSE) between the original image $X_t$ and the reconstructed image $\hat{X}_{t'}$ as the loss function. Formally, the training objective is:
\begin{equation}
\mathcal{L}_{\Theta_{\text{enc}}, \Theta_{\text{dec}}} = \mathbb{E}_{X_t \sim p_{X_t}} \left[ | X_t - \hat{X}_{t'} |_2^2 \right],
\end{equation}
where $\Theta_{\text{enc}}$ and $\Theta_{\text{dec}}$ denote the parameters of the encoder and decoder networks, respectively. 

\begin{figure}[t]
    \centering
    % \vspace{-0.5cm}
    \includegraphics[width=0.98\linewidth]{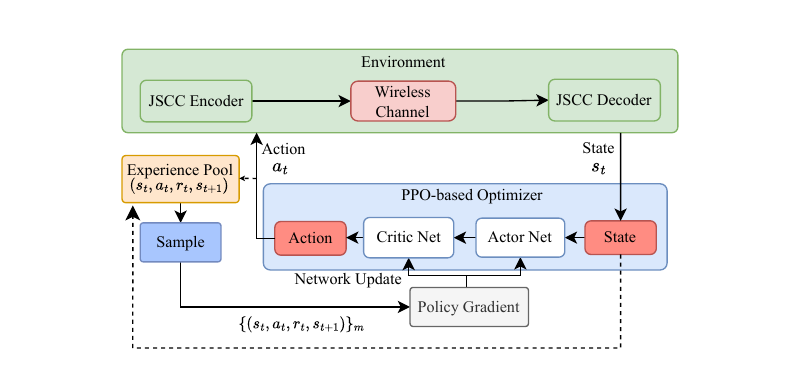}\vspace{-0.3cm}
    \caption{Architecture of proposed PPO-based optimizing algorithm.}
    \label{fig:drl} 
     \vspace{-0.6cm}
\end{figure}

\section{Timeliness-Aware Code Length Allocation Policy Optimization}

From the view of content-level semantics, empirical results show that increased bit rates tend to improve reconstructed image quality. However, an overly high bit rate necessitates larger packet sizes for data transmission, inevitably introducing latency and thereby impairing the system timeliness. Therefore, there exists a trade-off between image reconstruction quality and information timeliness. By treating the average PSNR as a constraint on reconstruction quality, the timeliness optimization problem can be formulated as follows:
\begin{align}
    \max_{\pi} & \lim_{T \to \infty} \frac{1}{T} \sum_{t=0}^{T} \text{VoI}(t)    \label{eq:problem_formulation}    \\
    \text{s.t. } & \lim_{T \to \infty} \frac{1}{T} \sum_{t=0}^{T} \text{PSNR}(X_t, \hat{X}_{t'}) \ge d_{\min} , \notag \\
    & \quad \pi(t) = \eta_t, \notag
\end{align}
where $\pi$ is the code length allocation policy. In this optimization problem, we aim to derive an effective code length allocation policy that strikes a balance between information timeliness and reconstruction quality under varying system conditions.

In order to achieve adaptive codeword length allocation balancing the timeliness requirements and reconstruction quality, we introduce a deep reinforcement learning (DRL) agent that learns to optimize the transmission policy, aiming to maximize information timeliness while guaranteeing an average reconstruction quality in Eq.~\ref{eq:problem_formulation}.

Different images exhibit varying sensitivity to transmission bitrate. While increasing the CBR generally enhances reconstruction quality, some images benefit more than others, and certain images already maintain high fidelity even at low bitrates. Such diverse reconstruction characteristics across different images motivate us to introduce an image transmission coefficient, which quantifies the image’s sensitivity to transmission. The transmission coefficient for image $X_t$ is defined as
\begin{equation}
\mu_t = \frac{\text{PSNR}^{\text{max}}_t - \text{PSNR}^{\text{min}}_t}{L_{\text{max}} - L_{\text{min}}} \times \text{PSNR}^{\text{max}}_t,
\end{equation}
where $\text{PSNR}^{\text{max}}_t$ and $\text{PSNR}^{\text{min}}_t$ represent the reconstruction quality under maximum code length $L_{\text{max}}$ and minimum code length $L_{\text{min}}$. This coefficient guides adaptive code length allocation, enabling the system to better balance transmission efficiency and reconstruction quality.

The proposed timeliness-aware semantic communication system can be formulated as a Markov Decision Process (MDP). This formulation enables sequential decision-making for code length allocation policy under dynamically environment conditions \cite{yang2024joint}.
At the transmission decision step, the agent observes a compact system state that reflects both the image semantic significance and the system’s recent transmission performance. The state $s_t \in \mathcal{S}$ at $t$ step is composed of three components:
\begin{equation}
s_t = [ \text{VoI}_t, \text{PSNR}_{t-1}, u(t), \mu_t ],
\end{equation}
where $\text{VoI}_t$ is current VoI, $\text{PSNR}_{t-1}$ is the reconstruction quality of last image $X_{t-1}$, $u(t)$ and $\mu_t$ are generation timestamp and transmission coefficient of image $X_t$. 

Based on the observed state $s_t$, the agent selects an action $a_t \in \mathcal{A}$ that determines the code length level for the current image. The action is defined as $a_t = \eta_t$, which directly controls the codeword length generated by the adaptive JSCC encoder.

\begin{algorithm}
\caption{Train the Timeliness-Aware Semantic Communication System}
\begin{algorithmic}[1]
\Input Training dataset $\mathcal{X}$, training steps $N_{1}, N_{2}$
\Ensure Optimized parameters $(\Theta_{\text{enc}}, \Theta_\text{dec}, \Theta_{\text{actor}}, \Theta_\text{critic})$
\Statex \textbf{Stage-1: Train the adaptive JSCC}
\State Initialize $(\Theta_{\text{enc}}, \Theta_\text{dec})$
\For{$i = 1$ \textbf{to} $N_{1}$}
    \State Sample $X_i \sim p_{X_i}, X_i\in\mathcal{X} $
    \State Randomly sample $\gamma_i, \eta_i$ in available set
    \State Apply adaptive JSCC to encode $X_i$ into codeword $\mathbf{x}_i$
    \State Transmit $\mathbf{x}_i$ over AWGN channel to obtain $\mathbf{y}_{i}$
    \State Decode $\mathbf{y}_{i}$ to obtain $\hat{X}_{i}$  
    \State Compute loss: $\mathcal{L} = \left[ | X_i - \hat{X}_{i} |_2^2 \right]$
    \State Update $(\Theta_{\text{enc}}, \Theta_\text{dec})$
\EndFor
\State Fix the parameters $(\Theta_{\text{enc}}, \Theta_\text{dec})$
\Statex \textbf{Stage-2: Train the DRL-based Optimizer}
\State Initialize $(\Theta_{\text{actor}}, \Theta_\text{critic})$
\For{$t = 1$ \textbf{to} $N_2$}
    \State Sample $X_t \sim p_{X_t}, X_t\in\mathcal{X} $
    \State Compute transmission coefficient $\mu_t$ for $X_t$
    \State Define state $s = (\text{VoI}(t),\text{PSNR}_{t-1},u(t),\mu_t)$
    \State Sample action $a_t \in \mathcal{A}$
    \State Encode image using JSCC under action $a_t$
    \State Transmit image and wait for successfully receiving
    \State Compute  $r_t = \text{VoI}(t)+\lambda\cdot( \text{PSNR}(X_t,\hat{X}_t)-d_\text{min})$
    \State Store $(s_t, a_t, r_t,s_{t+1})$ in rollout buffer
\If{rollout buffer is full}
    % \State Update policy $\Theta_\text{actor}$ via PPO surrogate loss
    % \State Update value network $\Theta_\text{cirtic}$ via regression loss
    \State Update $\Theta_{\text{actor}}$ by clipped PPO surrogate objective
    \State Update $\Theta_{\text{critic}}$ by value function loss
\EndIf
\EndFor
\State Fix the parameters $(\Theta_{\text{enc}}, \Theta_\text{dec}, \Theta_{\text{actor}}, \Theta_\text{critic})$
\end{algorithmic}
\label{alg:train}
\end{algorithm}

The goal of the DRL agent is to learn a policy $\pi$ that maps the current environment state to an optimal action, which can be expressed as $\pi: \mathcal{S} \rightarrow \mathcal{A}.$ The agent aims to learn this policy $\pi$ that can select the most appropriate codeword length for each image to maximize overall semantic timeliness while satisfying reconstruction quality constraints.

The design of the reward function plays a critical role in guiding the learning process. In our formulation, the reward at time step $t$ is defined as:
\begin{equation}
r_t = \text{VoI}(t) + \lambda \cdot \left( \text{PSNR}(X_t, \hat{X}_{t'}) - d_{\min} \right),
\end{equation}
where $\lambda \ge 0$ is a Lagrange multiplier that balances the trade-off between semantic timeliness and reconstruction quality. This reward drives the agent to balance information timeliness and reconstruction fidelity, encouraging timeliness-aware transmission while ensuring acceptable image quality. 

To optimize the code length allocation policy under the defined MDP framework, we adopt Proximal Policy Optimization (PPO) \cite{schulman2017proximal}, a stable and efficient reinforcement learning algorithm.
As shown in Fig.~\ref{fig:drl}, the PPO algorithm is implemented under an actor-critic framework, where the actor network $f_{\text{actor}}({s_t;\Theta_{\text{actor}}})$ maps the observed state to a probability distribution over discrete actions, and the critic network $f_{\text{critic}}(s_t;\Theta_{\text{critic}})$ estimates the state-value function to guide policy updates. 
During training, $\Theta_{\text{actor}}$ is updated by maximizing a clipped surrogate objective that ensures stable policy improvement, while $\Theta_{\text{critic}}$ is optimized by minimizing the mean-squared error between the predicted and target state values. In our framework, we follow the standard PPO formulation  \cite{schulman2017proximal} with clipped surrogate objectives and value function loss. The details of the training algorithm are summarized in Algorithm~\ref{alg:train}.

\section{Numerical Results}

In this section, we present experiments on the adaptive JSCC model and the code length allocation policy optimization, demonstrating their effectiveness in balancing the trade-off between reconstruction quality and image timeliness.

\subsection{Simulation Setup}

We evaluate the proposed timeliness-aware JSCC on the BuoyCAM dataset \cite{kohler2015scoop}, which contains 10,000 256×256 RGB images (8000 training, 2000 testing) with ocean backgrounds, ships, and varying weather conditions.
Both encoder and decoder adopt 6 Swin Transformer blocks. Training uses Adam (learning rate $1\times 10^{-4}$, batch size 8) for 200 epochs. PPO is trained for $1 \times 10^5$ steps with learning rate $1\times 10^{-4}$. And it uses two-layer fully connected actor/critic networks using Tanh activations and 64-dim latent states. To capture timeliness under limited bandwidth, transmission delay is modeled as $\Delta_{t} = K / \mathcal{R}$, where $K$ is the code length and $\mathcal{R}$ the baud rate.

% Both the encoder and decoder are implemented using 6 Swin Transformer blocks. As for the training configuration, we use the Adam optimizer with a learning rate of $1\times 10^{-4}$ and a batch size of 8. The model is trained for 200 epochs on the ocean monitoring dataset.

% The PPO algorithm is trained for a total of $1 \times 10^5$ environment interaction steps and at a learning rate of $1\times 10^{-4}$. Both the actor network and critic network adopt a two-layer fully connected architecture. Each network consists of two linear layers with a Tanh activation function layer applied in between. The input dimension matches the size of the state vector, and both layers map the features to a 64-dimensional latent representation. In addition, to account for the timeliness constraint under bandwidth-limited conditions, we explicitly consider the transmission delay, which is calculated as $\Delta_{t} = \frac{K}{\mathcal{R}}$, where $\mathcal{R}$ is the buad rate. 

To evaluate the effectiveness of the proposed adaptive JSCC scheme, we compare it with three baseline methods:
(1) JPEG+LDPC: a conventional JPEG compression followed by LDPC channel coding, representing a typical separation-based approach; 
(2) CNN-based JSCC: a CNN-based JSCC model that is a classic architecture of JSCC for image transmission\cite{bourtsoulatze2019deep}; 
(3) Non-adaptive SwinJSCC: a non-adaptive Swin-Transformer-based JSCC model that uses the same architecture as our method but without the adaptive encoding module; and
(4) DeepJSCC-l w/ Swin: an classic adaptive JSCC model but with a Swin Transformer backbone replacing the original CNN architecture\cite{kurka2021bandwidth}. 
As for evaluating the effectiveness of our PPO-based code length allocation policy optimizer, we compare our PPO-JSCC framework against a baseline consisting of JPEG compression combined with fixed quality parameters.

\begin{figure}[t]
    \centering
    \includegraphics[width=0.49\textwidth]{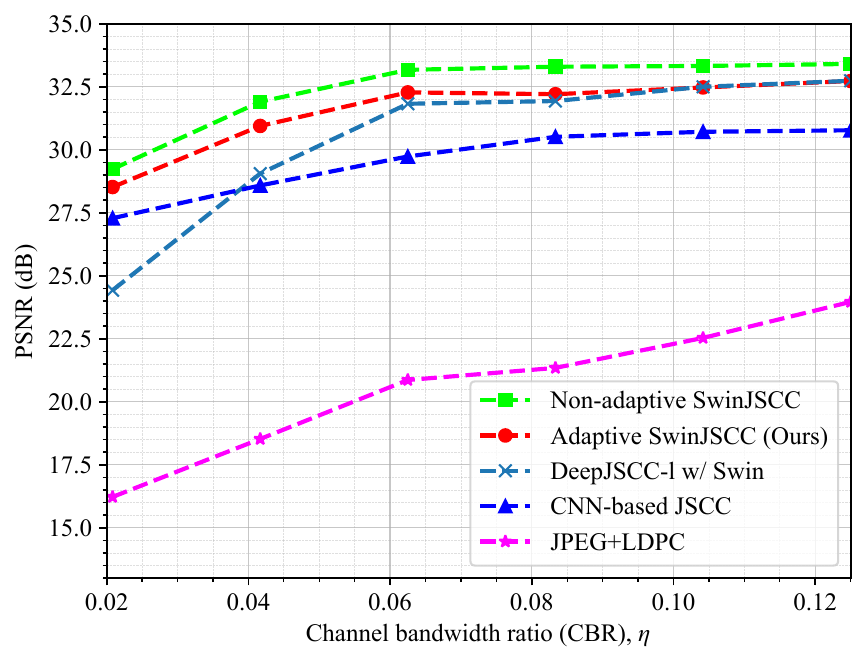} \vspace{-0.4cm}
    \caption{Average reconstruction PSNR against CBR with $\gamma=5$ dB.}
    \label{fig:cbr}
    \vspace{-0.6cm}
\end{figure}

\subsection{JSCC Performance Analysis}
Fig.~\ref{fig:cbr} illustrates the performance of different methods under varying channel bandwidth ratios. Our adaptive JSCC achieves an average reconstruction PSNR only $2.77\%$ lower than the non-adaptive JSCC baseline. Despite this small gap, the adaptive model offers greater flexibility and efficiency by supporting multiple codeword lengths with a single unified model. Compared with DeepJSCC-l w/ Swin, our method shows better performance under lower CBRs. In addition, our adaptive JSCC outperforms the CNN-based JSCC by $6.50\%$ in average reconstruction PSNR. It demonstrates the superiority of the Swin Transformer backbone in feature extraction and global context modeling. Besides, all JSCC-based methods significantly outperform the traditional JPEG+LDPC method, especially under low channel bandwidth ratios, highlighting the advantage of end-to-end learned semantic coding in bandwidth-limited scenarios. Moreover, all methods exhibit improved reconstruction quality as the CBR increases, which indicates that more transmission content leads to better preservation of content dimension semantic information.

\begin{figure}[t]
    \centering 
    \includegraphics[width=0.49\textwidth]{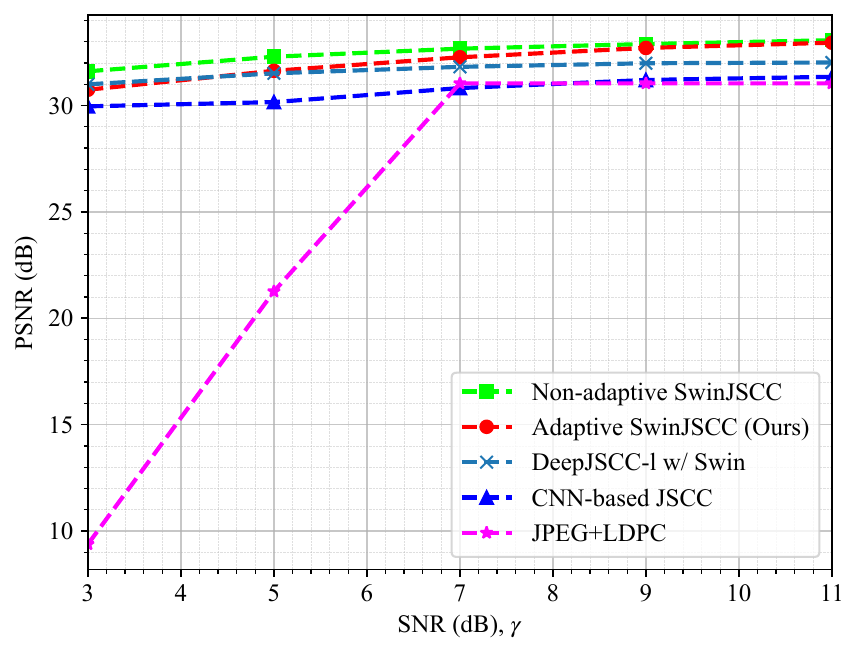} \vspace{-0.4cm}
    \caption{Average reconstruction PSNR against SNR with $\eta=\frac{3}{48}$.} 
    \label{fig:snr}  
    \vspace{-0.4cm}
\end{figure}

Fig.~\ref{fig:snr} illustrates the performance of different methods under varying channel SNRs. All methods show improved reconstruction quality with increasing SNR, reflecting the inherent dependence of transmission fidelity on channel conditions. Compared to the non-adaptive SwinJSCC baseline, our adaptive JSCC achieves an average reconstruction PSNR that is only $1.37\%$ lower across different SNR levels. Moreover, compared with DeepJSCC-l w/ Swin, our method attains superior reconstruction performance under favorable SNR conditions. In addition, our JSCC surpasses the CNN-based JSCC by $4.43\%$ in average reconstruction PSNR, further highlighting the advantages of the Swin Transformer backbone for feature representation. The adaptive JSCC model also significantly outperforms the conventional JPEG+LDPC scheme, achieving an average PSNR gain of $29.54\%$. Notably, the JPEG+LDPC method exhibits a pronounced cliff effect, resulting in a sharp degradation in reconstruction quality under low SNR conditions. Overall, the comparison between JSCC-based approaches and traditional separate coding confirms the effectiveness of semantic-aware JSCC strategies, particularly in resource-constrained scenarios.

The results presented in Fig.~\ref{fig:cbr} and Fig.~\ref{fig:snr} collectively validate the robustness of the proposed adaptive JSCC framework across varying code rates and channel conditions. This adaptability establishes a solid foundation for the subsequent optimization of transmission policies.

\subsection{Timeliness Performance Analysis}
\begin{figure}[t]
    \centering  
    \includegraphics[width=0.49\textwidth]{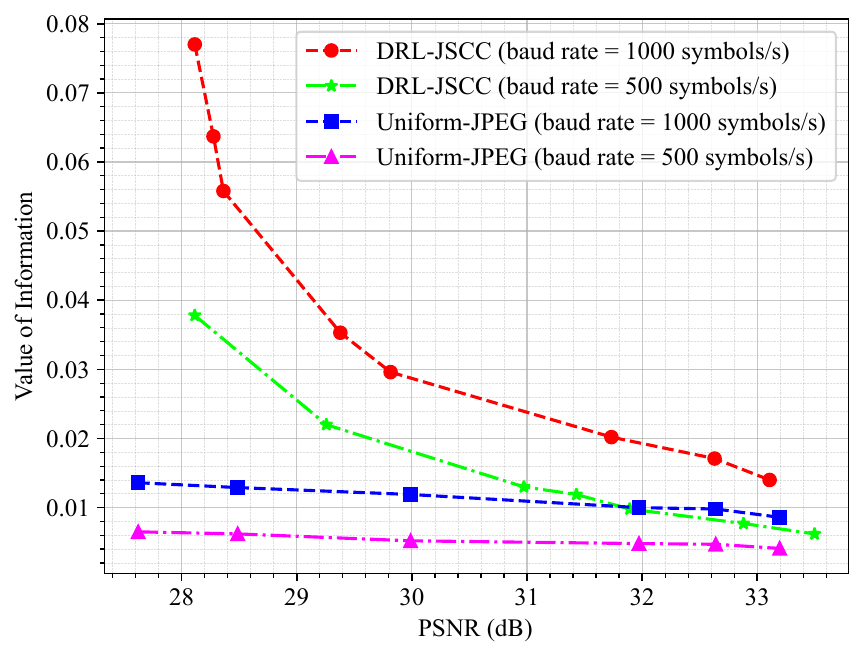} \vspace{-0.4cm}
    \caption{Average VoI against reconstruction PSNR constraints with $\rho=0.1$ and $\gamma = 7$ dB.} 
    \label{fig:voi}  
    \vspace{-0.6cm}
\end{figure}

Fig.~\ref{fig:voi} illustrates the performance comparison between the proposed DRL-JSCC method and the conventional Uniform-JPEG approach under different transmitted rates, in terms of the trade-off between average VoI and reconstruction PSNR.
Overall, as the reconstruction PSNR constraint increases, the timeliness performance of both methods declines. This reflects the inherent trade-off between information timeliness and reconstruction quality.
As shown in the figure, DRL-JSCC consistently outperforms Uniform-JPEG across all reconstruction PSNR constraints. Specifically, at a baud rate of $1000$ symbols/s, the average VoI of Uniform-JPEG decreases by $71.52\%$ compared with DRL-JSCC; while at $500$ symbols/s, the decrease is $66.10\%$.
These results suggest that DRL-JSCC consistently achieves significant timeliness advantages over Uniform-JPEG across different bandwidth conditions, demonstrating its effectiveness in preserving timeliness under varying transmission resources.
Clearly, the DRL-JSCC exhibits a higher timeliness performance than Uniform-JPEG when the reconstruction PSNR is low, and as the reconstruction PSNR constraint is increasing, the performance gap between the two methods gradually narrows. This is because ensuring higher reconstruction quality requires longer codewords, which reduces the flexibility of the DRL-based adaptive code length allocation policy. Consequently, the ability to optimize timeliness under strict PSNR constraints becomes limited.

\section{Conclusion}
In this paper, we propose a timeliness-aware JSCC framework for adaptive image transmission. The framework integrates adaptive JSCC with a deep reinforcement learning-based code length allocation policy to optimize the trade-off between semantic quality and information timeliness. The adaptive JSCC model utilizes CBR and SNR as side information to dynamically adjust the process of JSCC, while the code length allocation policy is optimized using the PPO algorithm to maximize the value of information under reconstruction quality constraints. Experimental results demonstrate that our approach significantly outperforms traditional separate coding schemes in terms of both reconstruction quality and timeliness, making it well-suited for timeliness-sensitive applications in wireless communication systems.

\bibliographystyle{IEEEtran}
\bibliography{refer}
\end{document}